\documentclass[aps,pra,showpacs,twocolumn]{revtex4}
\usepackage{graphics}
\usepackage{amsmath}
\usepackage{graphicx}
\usepackage{amsfonts}
\usepackage{amssymb}

\begin{document}
\title{Breaking entanglement-breaking by classical correlations}

\begin{abstract}
The inevitable interaction between quantum systems and environment induces
effects of decoherence which may be so strong as to destroy any initial
entanglement between the systems, a phenomenon known as \textquotedblleft
entanglement breaking\textquotedblright. Here we show the simplest examples
where the combination of two entanglement-breaking channels into a joint
correlated-noise environment reactivates the distribution of entanglement,
with classes of entangled states which are perfectly transmitted from a middle
station (Charlie) to two remote stations (Alice and Bob). Surprisingly, this
reactivation is induced by the presence of purely-classical correlations in
the joint environment, whose state is separable with zero discord. This
paradoxical effect is proven for quantum systems with Hilbert spaces of any
dimension, both finite and infinite.

\end{abstract}

\author{Stefano Pirandola}
\affiliation{Department of Computer Science, University of York, York YO10 5GH, United Kingdom}
\maketitle

\section{Introduction}

The distribution of entanglement is a central topic of investigation in the
quantum information community. Unfortunately, this distribution is also
fragile: Quantum systems inevitably interact with the external environment
whose decoherent action typically degrades their entanglement. The worst
scenario is when decoherence is so strong as to destroy any input
entanglement. Mathematically, this situation is represented by the concept of
entanglement-breaking channel~\cite{EBchannels,HolevoEB}. In general, a
quantum channel $\mathcal{E}$\ is entanglement-breaking when its local action
on one part of a bipartite state always results into a separable output state.
In other words, given two systems, $A$ and $B$, in an arbitrary bipartite
state $\rho_{AB}$, the output state $\rho_{AB^{\prime}}=(\mathcal{I}%
_{A}\otimes\mathcal{E}_{B})(\rho_{AB})$\ is always separable, where
$\mathcal{I}_{A}$ is the identity channel applied to system $A$ and
$\mathcal{E}_{B}$ is the entanglement-breaking channel applied to system $B$.

\begin{figure}[ptbh]
\vspace{-2.4cm}
\par
\begin{center}
\includegraphics[width=0.62\textwidth] {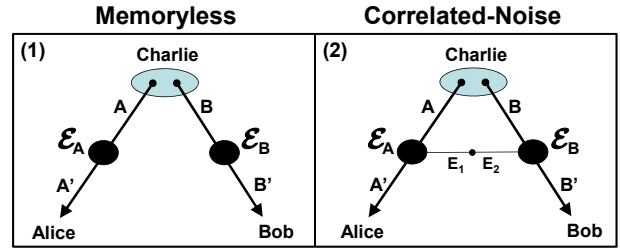}
\end{center}
\par
\vspace{-2.7cm}\caption{Symmetric scheme for entanglement distribution in
memoryless (1) and correlated-noise (2) environments. Charlie is the middle
station with an entangled state of systems $A$ and $B$. Alice and Bob are the
remote receiving stations. Charlie may choose to keep one system and send the
other to Alice or Bob (single transmission) or he may decide to send both the
systems to the remote stations (double transmission). In panel~(1) we consider
two memoryless entanglement-breaking channels, $\mathcal{E}_{A}$ and
$\mathcal{E}_{B}$. No entanglement can be distributed, neither via single
transmission (Charlie $\rightarrow$ Alice or Charlie $\rightarrow$Bob), nor
via double transmission (Charlie $\rightarrow$ Alice and Bob). In panel~(2) we
consider the presence of correlations between the two entanglement-breaking
channels. In this scenario, despite entanglement cannot be distributed by
single transmission (Charlie cannot be entangled with Alice or Bob), still it
can be distributed via the double transmission (so that Alice and Bob can
become entangled). This reactivation of entanglement distribution is possible
in the presence of classical correlations, i.e., for an environment (systems
$E_{1}$ and $E_{2}$) in a purely-classical state (separable, with
zero-discord).}%
\label{scenario}%
\end{figure}

Despite entanglement-breaking channels having been the subject of an intensive
study by the community, they have only been analyzed under Markovian
conditions of no memory. In other words, when the distribution involves two or
more systems, these systems are typically assumed to be perturbed in an
independent fashion, each of them subject to the same memoryless channel. For
instance, consider the scheme depicted in panel (1) of Fig.~\ref{scenario},
where a middle station (Charlie) has a bipartite system $AB$ in some entangled
state, but its communication lines with two remote parties (Alice and Bob) are
affected by entanglement-breaking channels $\mathcal{E}_{A}$ and
$\mathcal{E}_{B}$.

Under the assumption of memoryless channels, there is clearly no way to
distribute entanglement among any of the parties. Suppose that Charlie tries
to share entanglement with one of the remote parties by sending one of the two
systems while keeping the other (a scenario that we call \textquotedblleft%
1-system transmission\textquotedblright\ or just \textquotedblleft single
transmission\textquotedblright). For instance, Charlie may keep system $A$
while transmitting system $B$ to Bob. The action of $\mathcal{I}_{A}%
\otimes\mathcal{E}_{B}$ destroys the initial entanglement, so that systems $A$
(kept) and $B^{\prime}$ (transmitted) are separable. Symmetrically, the action
of $\mathcal{E}_{A}\otimes\mathcal{I}_{B}$ destroys the entanglement between
system $A^{\prime}$ (transmitted) and system $B$ (kept). Now suppose that
Charlie sends system $A$ to Alice and system $B$ to Bob (a scenario that we
call \textquotedblleft2-system transmission\textquotedblright\ or
\textquotedblleft double transmission\textquotedblright). Since the joint
action of the two\ channels is given by the tensor product $\mathcal{E}%
_{A}\otimes\mathcal{E}_{B}=(\mathcal{E}_{A}\otimes\mathcal{I}_{B}%
)(\mathcal{I}_{A}\otimes\mathcal{E}_{B}) $ quantum entanglement must
necessarily be destroyed. In other words, since we have 1-system
entanglement-breaking, then we must have 2-system entanglement-breaking.

In this paper we show that the previous implication is false when we introduce
correlations (i.e., a memory) between the two entanglement-breaking channels.
In other words, in the presence of correlated-noise environments, the double
transmission can successfully distribute entanglement despite the single
transmission being subject to entanglement-breaking. This is equivalent to say
that Charlie can transmit entanglement to Alice and Bob, despite not being
able to share any entanglement with them. The most interesting fact is that
the entanglement distribution via the double transmission is reactivated by
the presence of separable and purely-classical correlations in the joint
environment. This effect is proven for quantum systems with Hilbert spaces of
any dimension, both finite (qudits) and infinite (continuous variable
systems~\cite{BraREV2,RMP}).

In particular, here we are interested in showing the simplest examples of
correlated-noise environments which allow for a \textit{perfect distribution}
of entanglement via the double transmission (2-system entanglement-preserving)
while preventing any entanglement distribution via the single transmission
(1-system entanglement-breaking). These environments are constructed by using
the so-called twirling operators $U\otimes U$ (or $U\otimes U^{\ast}$). A
random unitary $U$ is applied to system $A$ and the same unitary (or its
conjugate) is applied to system $B$. While the local action of a random
unitary ($U\otimes I$ or $I\otimes U$) is entanglement-breaking, the
correlated action ($U\otimes U$) perfectly preserves specific classes of
entangled states (belonging to decoherent-free subspaces of the joint
correlated environment).

Since these \textquotedblleft twirling environments\textquotedblright\ are
based on local operations and classical communication (LOCC), they are
expected to introduce correlations which are separable (local) and, more
precisely, purely-classical. This feature can be checked by considering a
unitary dilation where an explicit system for the environment is introduced.
For the twirling environments, such a system can be described by a classical
state, i.e., separable with zero quantum discord.

Twirling environments can easily be constructed for quantum systems of any
dimension. In the case of two qudits (each with Hilbert space of dimension $d
$), the randomization of the twirling operators $U\otimes U$ (or $U\otimes
U^{\ast}$) is performed over the entire unitary group. In this case, it is
easy to identify states which are invariant under $U\otimes U$-twirling
(Werner states~\cite{Werner}) and $U\otimes U^{\ast}$-twirling (isotropic
states~\cite{HOROs}). In the specific case of qubits ($d=2$), we can restrict
the randomization to the basis of the Pauli operators, with the qubit Werner
state being invariant under Pauli twirling. Things are less trivial for
infinite dimension, in particular, for bosonic modes. In this case, we
restrict the twirling operators to the compact group of orthogonal symplectic
transformations, i.e., phase-space rotations. The bosonic twirling environment
so defined is non-Gaussian. It is interesting to see that, only for the
$U\otimes U^{\ast}$-twirling, we can identify Gaussian states which are
invariant and entangled: These are the two mode squeezed vacuum states, also
known as Einstein-Podolsky-Rosen (EPR) states~\cite{EPR,RMP}.

In terms of potential impact, our work opens new possibilities for
entanglement distribution in correlated-noise environments and memory
channels, where the presence of correlations can be exploited to recover from
entanglement breaking. Since entanglement recovery can be achieved by the
injection of separable purely-classical correlations, our work poses
fundamental questions on the intimate relations between local and nonlocal
correlations and, more generally, between classical and quantum correlations.

The paper is structured as follows. In Sec.~\ref{QubitSEC}, we consider qubits
and we show the simplest twirling environments (correlated Pauli
environments). In Sec.~\ref{TwirlSECTION}, we consider the general case of
qudits evolving in multidimensional twirling environments. These results can
also be applied to the case of qubits. Then, in Sec.~\ref{BOSsections}, we
consider bosonic systems and their evolution under non-Gaussian twirling
environments which are based on correlated phase-space rotations. Finally,
Sec.~\ref{SECconclusion} is for conclusion, with a number of
Appendices~\ref{DEPapp}, \ref{TwirlAVEapp}, \ref{HaarAVEsec}, \ref{DephaseAPP}
and~\ref{invGAUSapp} containing simple proofs and technical details.

\section{Qubits in correlated Pauli environments\label{QubitSEC}}

The most strikingly simple example can be constructed for qubits considering
the basis of the four Pauli operators $I$, $X$, $Y=iXZ$ and $Z$~\cite{Nielsen}%
. Given an arbitrary input state $\rho_{AB}$ of two qubits, we consider the
correlated Pauli channel
\begin{equation}
\mathcal{E}(\rho_{AB})=\sum_{k=0}^{3}p_{k}~(P_{k}\otimes P_{k})\rho_{AB}%
(P_{k}\otimes P_{k})^{\dagger}~,\label{qubitENV}%
\end{equation}
where $P_{k}\in\{I,X,Y,Z\}$ and $p_{k}$ are probabilities. This channel is
clearly simulated by random LOCCs. In fact, it is equivalent to extract a
random variable $K=\{k,p_{k}\}$, apply the Pauli unitary $P_{k}$\ to qubit $A
$, communicate $k$ and then apply the same unitary $P_{k}$\ to qubit $B$. It
is easy to check that the two-qubit channel of Eq.~(\ref{qubitENV}) does not
change if we replace the Pauli twirling operator $P_{k}\otimes P_{k}$ with the
alternative operator $P_{k}\otimes P_{k}^{\ast}$.

It is easy to write the unitary dilation of the correlated Pauli channel. It
is sufficient to introduce an environment composed by two systems $E_{1}$ and
$E_{2}$, each being a qudit with dimension $d=4$ and orthonormal basis
$\{\left\vert k\right\rangle \}_{k=0}^{3}$. Then, we can write%
\begin{equation}
\mathcal{E}(\rho_{AB})=\mathrm{Tr}_{E_{1}E_{2}}\left[  U(\rho_{AB}\otimes
\rho_{E_{1}E_{2}})U^{\dagger}\right]  ~,\label{PauliDilation}%
\end{equation}
where the environment is prepared in the correlated state%
\begin{equation}
\rho_{E_{1}E_{2}}=\sum_{k=0}^{3}p_{k}\left\vert k\right\rangle _{E_{1}%
}\left\langle k\right\vert \otimes\left\vert k\right\rangle _{E_{2}%
}\left\langle k\right\vert ~,\label{StateENV1}%
\end{equation}
and the unitary interaction $U=U_{E_{1}A}\otimes U_{E_{2}B}$ is a tensor
product of two control-Pauli unitaries,
\begin{equation}
U_{E_{1}A}=\sum_{k=0}^{3}\left\vert k\right\rangle _{E_{1}}\left\langle
k\right\vert \otimes P_{k},~U_{E_{2}B}=\sum_{k=0}^{3}\left\vert k\right\rangle
_{E_{2}}\left\langle k\right\vert \otimes P_{k}.\label{Cunitaries}%
\end{equation}
As evident from Eq.~(\ref{StateENV1}), the state of the environment is
separable, which means that only separable (i.e., local-type) correlations are
injected into the travelling qubits. More precisely, since $\rho_{E_{1}E_{2}}$
is expressed as a convex combination of orthogonal projectors, it is a
purely-classical state, i.e., a state with zero quantum discord~\cite{RMPdis}.
As a result, this environment contains correlations which are not only local
but also purely classical.

Now we show the conditions under which the correlated Pauli environment is
simultaneously one-qubit entanglement-breaking and two-qubit entanglement
preserving. We start by considering the transmission of one qubit only, e.g.,
qubit $A$. In this case, Eq.~(\ref{qubitENV}) reduces to a depolarizing
channel
\begin{gather}
(\mathcal{E}_{A}\otimes\mathcal{I}_{B})(\rho_{AB})=\sum_{k=0}^{3}p_{k}%
(P_{k}\otimes I)\rho_{AB}(P_{k}^{\dagger}\otimes I)\nonumber\\
=p_{0}\rho_{AB}+\sum_{k=1}^{3}p_{k}(P_{k}\otimes I)\rho_{AB}(P_{k}^{\dagger
}\otimes I).\label{qubitDEPO}%
\end{gather}
It is easy to show that $\mathcal{E}_{A}$\ is entanglement-breaking when
$p_{k}\leq1/2$ for any $k$ (see Appendix~\ref{DEPapp}\ for a simple proof). A
particular choice can be $p_{0}\leq1/2$ and $p_{1}=p_{2}=p_{3}=(1-p_{0})/3$ as
for instance used in Ref.~\cite{Sacchi}.

Assuming the condition of one-qubit entanglement-breaking\ ($p_{k}\leq1/2$),
Charlie is clearly not able to share any entanglement with Alice or Bob. Can
he still distribute entanglement to Alice and Bob? Yes, this is possible
because we can identify a class of entangled states which are invariant under
the action of the correlated map~(\ref{qubitENV}). This class is simply given
by the Werner states%
\begin{equation}
\rho_{AB}(\gamma):=(1-\gamma)\frac{I_{AB}}{4}+\gamma\left\vert -\right\rangle
_{AB}\left\langle -\right\vert ~,\label{2qubitWERNER}%
\end{equation}
with parameter $-1/3\leq\gamma\leq1$, where $I_{AB}/4$ is the maximally-mixed
state and
\begin{equation}
\left\vert -\right\rangle _{AB}=\frac{1}{\sqrt{2}}\left(  \left\vert
0\right\rangle _{A}\left\vert 1\right\rangle _{B}-\left\vert 1\right\rangle
_{A}\left\vert 0\right\rangle _{B}\right)
\end{equation}
is the maximally-entangled (singlet) state. For $\gamma>1/3$, the two-qubit
state $\rho_{AB}(\gamma)$ is known to be entangled and distillable. Since
qubit Werner states are invariant under any twirling operator $U\otimes U$,
i.e.,%
\begin{equation}
(U\otimes U)\rho_{AB}(\gamma)(U\otimes U)^{\dagger}=\rho_{AB}(\gamma)~,
\end{equation}
for any unitary $U$, they are fixed points of the correlated
map~(\ref{qubitENV}) or, in other words, they represent a decoherence-free
subspace of the correlated Pauli environment. Thus, if Charlie sends Werner
states with $\gamma>1/3$, these entangled states are perfectly distributed to
Alice and Bob (two-qubit entanglement preserving).

\section{Qudits in multidimensional twirling environments\label{TwirlSECTION}}

In this section we consider quantum systems whose Hilbert space has finite
dimension $d$, i.e., qudits (in particular, we have qubits for $d=2$). For
these systems, we can easily construct classically-correlated environments
which are simultaneously 1-qudit entanglement-breaking and 2-qudit entanglement-preserving.

Consider two qudits, $A$ and $B$, with same dimension $d$ and prepared in a
bipartite state $\rho_{AB}$. Then, we call $U\otimes U$ twirling channel the
following completely positive trace-preserving map%
\begin{equation}
\mathcal{E}_{UU}(\rho_{AB})=\int_{\mathcal{U}(d)}dU~(U\otimes U)\rho
_{AB}(U\otimes U)^{\dagger}~,\label{GENtwMAP}%
\end{equation}
where the integral is over the entire unitary group $\mathcal{U}(d)$ acting on
the $d$-dimensional Hilbert space, and $dU$ is the Haar measure. This channel
is clearly realizable by random LOCCs. In fact, it is equivalent to choose a
random unitary $U$, apply it to the local system $A$, and then apply the same
$U$ to the other local system $B$ (where the coordination of the two local
unitaries is mediated by classical communication). Similarly, we can define a
$U\otimes U^{\ast}$ twirling channel, by replacing the twirling operator
$U\otimes U$ with the alternative twirling $U\otimes U^{\ast}$ in the
definition of Eq.~(\ref{GENtwMAP}). Compactly, we refer to the $U\otimes V$
twirling channel%
\begin{equation}
\mathcal{E}_{UV}(\rho_{AB})=\int_{\mathcal{U}(d)}dU~(U\otimes V)\rho
_{AB}(U\otimes V)^{\dagger}~,\label{genTWIRL}%
\end{equation}
where $V=U$ or $V=U^{\ast}$.

In order to study its correlation properties we dilate this channel into an
environment. Note that, since we have an integral in Eq.~(\ref{genTWIRL}), the
dilation of the channel seems to involve the introduction of continuous
variable systems. In fact, the unitary group $\mathcal{U}(d)$ is described by
$d^{2}$ real parameters~\cite{UnitaryPAR}, which means that we would need to
employ $d^{2}$ continuous variable systems for the environment. Actually, this
continuous dilation is not necessary, since can always replace the previous
Haar integral with a discrete sum over a finite number of suitably-chosen
unitaries. In fact, any twirling channel~(\ref{genTWIRL}) can be written as%
\begin{equation}
\mathcal{E}_{UV}(\rho_{AB})=\frac{1}{K}\sum_{k=0}^{K-1}(U_{k}\otimes
V_{k})\rho_{AB}(U_{k}\otimes V_{k})^{\dagger}~,\label{design1}%
\end{equation}
where $U_{k}$ belongs to the set of unitary 2-design $\mathcal{D}%
$~\cite{Designs,Des2} and $V_{k}=U_{k}$ or $V_{k}=U_{k}^{\ast}$. The set
$\mathcal{D}$ has a finite number of elements which depends of the dimension
of the Hilbert space $K=K(d)$ (too see how the cardinality $K$ scales with the
dimension $d$, see for instance Ref.~\cite{Gross}). The proof of the
equivalence between Eqs.~(\ref{genTWIRL}) and~(\ref{design1}) can be found in
Ref.~\cite{Designs} for the $U\otimes U$ twirling channel. See
Appendix~\ref{TwirlAVEapp}\ for a simple extension of the proof to the other
$U\otimes U^{\ast}$ twirling channel. Note that, in the case of qubits
($d=2$), an example of unitary 2-design is provided by the Clifford
group~\cite{Designs,Clifford3}, which is the normalizer of the Pauli group and
typically employed in quantum error correction~\cite{Clifford1,Clifford2}.

Now using the unitary 2-design, we can dilate the twirling channel into an
environment made by finite-dimensional systems, i.e., two larger qudits
$E_{1}$ and $E_{2}$, each with dimension $K$ and orthonormal basis
$\{\left\vert k\right\rangle \}_{k=0}^{K-1}$. In fact, we can write%
\begin{equation}
\mathcal{E}_{UV}(\rho_{AB})=\mathrm{Tr}_{E_{1}E_{2}}\left[  U(\rho_{AB}%
\otimes\rho_{E_{1}E_{2}})U^{\dagger}\right]  ~,
\end{equation}
where the environment is prepared in the uniformly correlated state%
\begin{equation}
\rho_{E_{1}E_{2}}=\frac{1}{K}\sum_{k=0}^{K-1}\left\vert k\right\rangle
_{E_{1}}\left\langle k\right\vert \otimes\left\vert k\right\rangle _{E_{2}%
}\left\langle k\right\vert ~,\label{StateENV2}%
\end{equation}
and the unitary interaction $U=U_{E_{1}A}\otimes U_{E_{2}B}$ is a tensor
product of two control-unitaries
\begin{equation}
U_{E_{1}A}=\sum_{k=0}^{K-1}\left\vert k\right\rangle _{E_{1}}\left\langle
k\right\vert \otimes U_{k},~U_{E_{2}B}=\sum_{k=0}^{K-1}\left\vert
k\right\rangle _{E_{2}}\left\langle k\right\vert \otimes V_{k}%
,\label{Cunitaries2}%
\end{equation}
where $U_{k}\in\mathcal{D}$ and $V_{k}=U_{k}$ or $V_{k}=U_{k}^{\ast}$. As
evident from Eq.~(\ref{StateENV2}), the state of the environment is separable
and purely-classical (zero discord). In other words, twirling environments
only contain purely-classical correlations.

Once we have characterized the correlation properties of these environments,
we show the conditions under which they are simultaneously one-qudit
entanglement-breaking and two-qudit entanglement preserving. First of all, let
us explicitly show that one-qudit transmission is always subject to
entanglement breaking. Suppose that only qudit $A$ is transmitted by Charlie.
Then, for any input state $\rho_{AB}$, the output state of Alice and Charlie
is given by%
\begin{align}
(\mathcal{E}_{A}\otimes\mathcal{I}_{B})(\rho_{AB})  &  =\nonumber\\
\int_{\mathcal{U}(d)}dU~(U\otimes I)\rho_{AB}(U^{\dagger}\otimes I)  &
=\frac{1}{d}I\otimes\mathrm{Tr}_{A}(\rho_{AB})~,\label{depola2}%
\end{align}
which means that the random map $U\otimes I$ represents an
entanglement-breaking channel (similar result holds for the other random map
$I\otimes V$ involved in the transmission of the other qudit $B$). As shown in
Appendix~\ref{HaarAVEsec}, the proof of Eq.~(\ref{depola2}) is a simple
application of the identity
\begin{equation}
\left\langle O\right\rangle _{U}:=\int_{\mathcal{U}(d)}dU~UOU^{\dagger}%
=\frac{\mathrm{Tr}(O)}{d}I~,\label{twirlingID}%
\end{equation}
which is the Haar average of a linear operator $O$. In turn,
Eq.~(\ref{twirlingID}) is a simple consequence of Schur's lemma and the
invariance of the Haar measure~\cite{Chiri}.

The next step is to consider two-qudit transmission from Charlie to Alice and
Bob. In this case, we search for entangled states which are preserved by the
correlated action of the twirling environment. Luckily, we can easily find
states which are invariant under the action of the twirling operator $U\otimes
V$, i.e.,
\begin{equation}
(U\otimes V)\rho_{AB}(U\otimes V)^{\dagger}=\rho_{AB}~.\label{TwirlINVA}%
\end{equation}
Thanks to this invariance, such states are fixed points of the $U\otimes V$
twirling channel of Eq.~(\ref{genTWIRL}).

In the specific case of $V=U$, it is well known that the unique solution of
Eq.~(\ref{TwirlINVA}) is provided by the multidimensional Werner states, which
are themselves defined as those states invariant under $U\otimes U$
twirling~\cite{Werner}. Given two isodimensional qudits, $A$ and $B$, their
$d\times d$ Werner state is a one-parameter class defined
by~\cite{Werner,Synak}%
\begin{equation}
\rho_{AB}(\mu):=\frac{1}{d^{2}+d\mu}(I_{AB}+\mu V)\label{multiWerner}%
\end{equation}
where $-1\leq\mu\leq1$ and $V$ is the unitary flip operator $V\left\vert
\varphi\right\rangle _{A}\otimes\left\vert \psi\right\rangle _{B}=\left\vert
\psi\right\rangle _{A}\otimes\left\vert \varphi\right\rangle _{B}$. This state
is known to be entangled for $\mu<-d^{-1}$. Thus, if Charlie has a Werner
state $\rho_{AB}(\mu)$ with suitable $\mu$ (in the entanglement regime), he is
able to perfectly transmit his state to Alice and Bob, who can then share and
distill entanglement. This entanglement distribution from Charlie to Alice and
Bob is possible, despite Charlie cannot share any entanglement with the remote
parties due to the entanglement breaking condition of Eq.~(\ref{depola2}).

Coming back to Eq. (\ref{TwirlINVA}), one can find a similar solution for
$V=U^{\ast}$. In fact, as shown in Ref.~\cite{HOROs}, there exist states which
are invariant under $U\otimes U^{\ast}$-twirling. These are called isotropic
states, and they are defined by the one-parameter class~\cite{HOROs}%
\begin{equation}
\rho_{AB}(\gamma):=(1-\gamma)\frac{I_{AB}}{d^{2}}+\gamma\left\vert
\psi\right\rangle _{AB}\left\langle \psi\right\vert ~,\label{Isostates}%
\end{equation}
where the maximally mixed state $d^{-2}I_{AB}$\ and the maximally entangled
state
\begin{equation}
\left\vert \psi\right\rangle _{AB}=\frac{1}{\sqrt{d}}\sum_{k=0}^{d-1}%
\left\vert k\right\rangle _{A}\left\vert k\right\rangle _{B}%
~,\label{maxENTstate}%
\end{equation}
are combined with parameter $-(d^{2}-1)^{-1}\leq\gamma\leq1$. In general, it
is entangled and distillable for $\gamma>(1+d)^{-1}$. Thus, if Charlie has an
isotropic state $\rho_{AB}(\gamma)$ with suitable $\gamma$ (i.e., in the
entanglement regime), he is able to perfectly transmit this state to Alice and
Bob, who therefore can share and distill entanglement. Again, this is possible
despite the transmission of a single qudit is subject to an
entanglement-breaking channel, as expressed by Eq.~(\ref{depola2}).

It is clear that these results can be specialized to the case of qubits ($d=2
$). For qubits, the classes of multidimensional Werner states of
Eq.~(\ref{multiWerner}) and isotropic states of Eq.~(\ref{Isostates}) coincide
up to a local unitary~\cite{HOROs}. Multidimensional Werner states reduce to
the qubit Werner state of Eq.~(\ref{2qubitWERNER}) which is $U\otimes
U$-invariant. In fact, for $d=2$, we have $(I_{AB}+\mu V)=(\mu+1)I_{AB}%
-2\mu\left\vert -\right\rangle _{AB}\left\langle -\right\vert $ which,
replaced in Eq.~(\ref{multiWerner}), gives the state of
Eq.~(\ref{2qubitWERNER}) by setting $\gamma=-\mu/(2+\mu)$.

On the other hand, isotropic states reduce to Eq.~(\ref{2qubitWERNER}),
proviso that the singlet $\left\vert -\right\rangle _{AB}$ is replaced by the
triplet
\begin{equation}
\left\vert +\right\rangle _{AB}=\frac{1}{\sqrt{2}}\left(  \left\vert
0\right\rangle _{A}\left\vert 0\right\rangle _{B}+\left\vert 1\right\rangle
_{A}\left\vert 1\right\rangle _{B}\right)  ~.\label{tripletSTATE}%
\end{equation}
This state is $U\otimes U^{\ast}$-invariant and known as Werner-like state.
Since singlet and triplet states are connected by a local unitary, there is no
difference in the quantum correlations which are contained in the state
according to the two definitions. More generally, in a Werner-like state, any
maximally entangled state could be considered in the place of the singlet.

\section{Bosonic twirling environments\label{BOSsections}}

Here we extend the analysis to the case of continuous variable systems, i.e.,
quantum systems with infinite dimensional Hilbert spaces ($d=\infty$). In
particular, we consider the case of two bosonic modes of the electromagnetic
field. The simplest generalization of the notion of twirling environment
involves the use of rotations in the phase space. Given a single bosonic mode
with number operator $\hat{n}$, the rotation operator is defined as $\hat
{R}_{\theta}=\exp(-i\theta\hat{n})$. In the phase space, the action of this
operator is described by the well-known rotation matrix
\begin{equation}
\mathbf{R}_{\theta}=\left(
\begin{array}
[c]{cc}%
\cos\theta & \sin\theta\\
-\sin\theta & \cos\theta
\end{array}
\right)  ~.
\end{equation}
In terms of the second-order statistical moments, we have that the covariance
matrix (CM) $\mathbf{V}$ of the input mode is transformed via the congruence
\begin{equation}
\mathbf{V}\rightarrow\mathbf{R}_{\theta}\mathbf{VR}_{\theta}{}^{T}~.
\end{equation}

Now, given an input state $\rho_{AB}$ of two modes, $A$ and $B$, we can
synchronize two random rotations and define the bosonic twirling channel%
\begin{equation}
\mathcal{E}_{\theta\theta^{\prime}}(\rho_{AB})=\int\frac{d\theta}{2\pi}%
~(\hat{R}_{\theta}\otimes\hat{R}_{\theta^{\prime}})\rho_{AB}(\hat{R}_{\theta
}\otimes\hat{R}_{\theta^{\prime}})^{\dagger},\label{ROTchannel}%
\end{equation}
where $\theta^{\prime}=\theta$ or $\theta^{\prime}=-\theta$. This is a
non-Gaussian channel, since the twirling operator $\hat{R}_{\theta}\otimes
\hat{R}_{\theta^{\prime}}$, despite Gaussian, is not averaged using a Gaussian
distribution but a uniform one. It is clearly based on random LOCCs, since
random rotations are locally applied to each bosonic mode and they are
correlated via classical communication.

It is interesting that the unitary dilation of this channel can be restricted
to a finite-dimensional environment. The reason is because the single-mode
rotation operator $\hat{R}_{\theta}$ belongs to the compact subgroup of the
orthogonal symplectic transformations $\mathcal{K}(2)=\mathcal{S}%
p(2)\cap\mathcal{O}(2)$ (which are passive Gaussian unitaries, i.e., unitary
transformations preserving both the Gaussian statistics and the energy of the
state). Then, $\mathcal{K}(2)$ is isomorphic to the unitary group
$\mathcal{U}(1)$, which is the multiplicative group composed by all complex
numbers with module 1, also known as the circle group. For this reason, a
unitary 2-design $\mathcal{D}\subset\mathcal{U}(1)$ can be mapped into a
unitary 2-design for $\mathcal{K}(2)$~\cite{Gross}. As a result, we can write
\begin{equation}
\mathcal{E}_{\theta\theta^{\prime}}(\rho_{AB})=\frac{1}{K}\sum_{k=0}%
^{K-1}(\hat{R}_{\theta_{k}}\otimes\hat{R}_{\theta_{k}^{\prime}})\rho_{AB}%
(\hat{R}_{\theta_{k}}\otimes\hat{R}_{\theta_{k}^{\prime}})^{\dagger},
\end{equation}
for a suitable set of angles $\{\theta_{0},\ldots,\theta_{K-1}\}$\ and where
$\theta_{k}^{\prime}=\theta_{k}$ or $\theta_{k}^{\prime}=-\theta_{k}$. In this
form, the channel is manifestly non Gaussian. Then, as before, it can be
represented using two environmental qudits, $E_{1}$ and $E_{2}$, prepared in a
correlated state $\rho_{E_{1}E_{2}}$ as in Eq.~(\ref{StateENV2}) and
interacting with the two bosonic modes via two control-unitaries as in
Eq.~(\ref{Cunitaries2}), where now the unitaries are rotations in the phase
space. The environmental state is not only separable but also purely-classical
(zero discord), which means that only classical correlations are injected by
the bosonic twirling environment.

It is easy to show that one-mode transmission is always subject to
entanglement-breaking in this environment. For instance, if mode $A$ is
transmitted from Charlie to Alice, then the output state
\begin{equation}
\rho_{A^{\prime}B}=\int\frac{d\theta}{2\pi}~(\hat{R}_{\theta}\otimes
I)\rho_{AB}(\hat{R}_{\theta}\otimes I)^{\dagger}\label{UNIdepha}%
\end{equation}
is separable, no matter what the input state $\rho_{AB}$ is. Indeed it is easy
to prove that a uniformly dephasing channel as that of Eq.~(\ref{UNIdepha}) is
entanglement-breaking. For the sake of completeness, we give this simple proof
in Appendix~\ref{DephaseAPP}.

The next step is to find two-mode states which are invariant under correlated
phase rotations%
\begin{equation}
(\hat{R}_{\theta}\otimes\hat{R}_{\theta^{\prime}})\rho_{AB}(\hat{R}_{\theta
}\otimes\hat{R}_{\theta^{\prime}})^{\dagger}=\rho_{AB}~,\label{InvROT}%
\end{equation}
so that they are perfectly transmitted by the bosonic twirling environment%
\begin{equation}
\mathcal{E}_{\theta\theta^{\prime}}(\rho_{AB})=\rho_{AB}~.
\end{equation}
For simplicity we restrict our search to zero-mean Gaussian states, therefore
completely characterized by their CMs. Let us call $\mathbf{V}_{AB} $ the
CM\ of an input Gaussian state $\rho_{AB}$. Then, finding a solution of
Eq.~(\ref{InvROT}) is equivalent to solve
\begin{equation}
(\mathbf{R}_{\theta}\mathbf{\oplus R}_{\theta^{\prime}})\mathbf{V}%
_{AB}(\mathbf{R}_{\theta}\mathbf{\oplus R}_{\theta^{\prime}})^{T}%
=\mathbf{V}_{AB}~.\label{InvROT2}%
\end{equation}
Depending on the type of environment, i.e., $\theta^{\prime}=\theta$ (perfect
correlation)\ or $\theta^{\prime}=-\theta$ (perfect anti-correlation), we have
two different classes of invariant Gaussian states.

Unfortunately, in the case of the $\hat{R}_{\theta}\otimes\hat{R}_{\theta} $
environment, the invariant Gaussian states are separable. In fact, it is easy
to check that, for $\theta^{\prime}=\theta$ and arbitrary $\theta$, the unique
solution of Eq. (\ref{InvROT2}) is given by the quasi-normal form%
\begin{equation}
\mathbf{V}_{AB}:=\left(
\begin{array}
[c]{cc}%
\mathbf{A} & \mathbf{C}\\
\mathbf{C}^{T} & \mathbf{B}%
\end{array}
\right)  =\left(
\begin{array}
[c]{cccc}%
\alpha &  & \omega & \varphi\\
& \alpha & -\varphi & \omega\\
\omega & -\varphi & \beta & \\
\varphi & \omega &  & \beta
\end{array}
\right)  ~,\label{CM_appfinite}%
\end{equation}
with $\alpha,\beta\geq1$ and $\omega,\delta$ are real numbers (which must
satisfy a set of bona-fide conditions in order to make the previous matrix a
quantum CM~\cite{TwomodePRA}). It is easy to check that the previous CM can
only describe separable Gaussian states. See Appendix~\ref{invGAUSapp} to see
how to derive the CM of Eq.~(\ref{CM_appfinite}) and check its separability.

Thus, despite there exist two-mode Gaussian states $\rho_{AB}$ which are
invariant under perfectly-correlated phase rotations $\hat{R}_{\theta}%
\otimes\hat{R}_{\theta}$, these states must be separable. This negative result
can be generalized: No entangled Gaussian state is invariant under twirlings
of the form $U\otimes U$, with $U$ Gaussian unitary (apart from the trivial
case $U=\pm I$). In fact, suppose that we have an input Gaussian state
$\rho_{AB}$ with mean value $\mathbf{\bar{x}}_{AB}$ and covariance CM
$\mathbf{V}_{AB}$. The action of two Gaussian unitaries $U\otimes U$ on
$\rho_{AB}$ corresponds to apply two identical displacements to its mean
value
\begin{equation}
\mathbf{\bar{x}}_{AB}\rightarrow\mathbf{\bar{x}}_{AB}+(\mathbf{d}%
,\mathbf{d})^{T}~,
\end{equation}
and two identical symplectic matrices to its CM
\begin{equation}
\mathbf{V}_{AB}\rightarrow(\mathbf{S\oplus S})\mathbf{V}_{AB}(\mathbf{S\oplus
S})^{T}~.
\end{equation}
In general, there is clearly no possibility to find an invariant Gaussian
state, since any nonzero displacement $\mathbf{d}$ maps the input state into a
different output state. We then restrict the search to considering canonical
Gaussian unitaries ($\mathbf{d=0}$) which are one-to-one with the symplectic
transformations. Unfortunately, this is still not the case. According to
Euler's decomposition~\cite{RMP}, any single-mode symplectic transformation
$\mathbf{S}$ is generally decomposed into a orthogonal rotation $\mathbf{R}%
_{\theta}$, a single-mode squeezing $\mathbf{S}_{r}$, and another orthogonal
rotation $\mathbf{R}_{\omega}$, where $\theta,\omega$ are angles and $r\geq1$
is a squeezing parameter. As long as squeezing is present ($r>1$), we have
that the trace of the CM changes (physically this corresponds to injecting
energy into the state). As a result, the CM and, therefore, the Gaussian state
must change. Thus, the only possibility is to find Gaussian states which are
invariant under rotations only (which are passive transformations, i.e.,
preserving the trace). However, we have already seen that, despite they exist,
these Gaussian states must be separable.

Luckily, the scenario is completely different when we consider the other type
of environment. We can easily find entangled Gaussian states which are
invariant under anti-correlated phase rotations $\hat{R}_{\theta}\otimes
\hat{R}_{-\theta}$. One can check that, for $\theta^{\prime}=-\theta$ and
arbitrary $\theta$, the unique solution of Eq. (\ref{InvROT2}) is given by the
CM%
\begin{equation}
\mathbf{V}_{AB}:=\left(
\begin{array}
[c]{cc}%
\mu\mathbf{I} & \sqrt{\mu^{2}-1}\mathbf{Z}\\
\sqrt{\mu^{2}-1}\mathbf{Z} & \mu\mathbf{I}%
\end{array}
\right)  ~,
\end{equation}
where $\mu\geq1$, $\mathbf{I}$ is the $2\times2$ identity matrix, and
$\mathbf{Z}=\mathrm{diag}(1,-1)$ is the reflection matrix. This is the CM of a
two-mode squeezed vacuum state, i.e., an EPR state~\cite{RMP}. Thus, in the
presence of a $\hat{R}_{\theta}\otimes\hat{R}_{-\theta}$ twirling channel,
despite Charlie is not able to share any entanglement with Alice or Bob
(one-mode entanglement breaking), he is still able to distribute entanglement
to them by transmitting EPR states perfectly (two-mode entanglement
preserving). As discussed in Ref.~\cite{PIRarxiv}, a more general class of
$\hat{R}_{\theta}\otimes\hat{R}_{-\theta}$ invariant states is given by the
continuous variable Werner states (which are generally non-Gaussian, and they
are constructed by mixing an EPR state with a tensor product of thermal states).

\section{Conclusion and discussion~\label{SECconclusion}}

In conclusion, we have investigated the distribution of entanglement in the
presence of correlated-noise environments, in particular, twirling
environments. We have considered quantum systems with Hilbert spaces of any
dimension, i.e., qubits, qudits, and bosonic systems. We have assumed the
condition of one-system entanglement breaking, meaning that the transmission
of a single system, e.g., from Charlie to Alice, cannot distribute any
entanglement. Despite this, we have shown that the distribution of
entanglement is still possible when we consider the double transmission, i.e.,
from Charlie to both Alice and Bob. In particular, we can identify classes of
entangled states which are invariant under the action of the composite
environment, which means that the entanglement is perfectly preserved in the
two-system transmission.

This effect must be ascribed to the correlations which are injected into the
travelling systems by the twirling environment. Interestingly, these
environmental noise-correlations are separable, i.e., local-type, and more
precisely purely-classical, since no quantum discord can be found in the state
of the environment. The fact that separability, and in particular,
classicality, can be exploited to recover from entanglement-breaking is a
paradoxical behavior which poses fundamental questions on the intimate
relations between classical and quantum correlations.

It is important to note that, despite twirling environments being very simple
examples, they are also quite artificial. As a matter of fact, such kind of
perfectly correlated quantum operations are typically used by Alice and Bob in
protocols of entanglement distillation (where the random twirling $U\otimes U$
is applied to transform bipartite states into Werner states, which are then
distilled into maximally entangled states).

Luckily, we can also prove that the reactivation of entanglement distribution
occurs in more realistic scenarios. An important non-trivial example can be
given for bosonic systems. Ref.~\cite{PIRarxiv} considers a realistic model of
correlated-noise Gaussian environment, which generalizes the standard
memoryless thermal-loss environment. In this Gaussian environment, the
presence of weak separable correlations is sufficient to reactivate the
distribution of entanglement from Charlie to Alice and Bob, despite the
thermal noise present in the single transmissions (Charlie-Alice or
Charlie-Bob) being entanglement-breaking. This case is non-trivial also
because we cannot identify any decoherent-free subspace of entangled Gaussian
states, which means that Gaussian entanglement cannot be preserved in the
double transmission. Despite perfect preservation of entanglement being not
possible, Charlie can still use the double transmission to distribute a
distillable\ amount of entanglement to the remote parties by sending EPR
states with sufficiently large squeezing.

\section{Acknowledgements}

This work has been supported by EPSRC under the research grant HIPERCOM
(EP/J00796X/1). Special thanks of the author are for S. L. Braunstein, M.
Paternostro and C. Ottaviani. The author would also like to thank (in random
order) P. Horodecki, O. Oreshkov, A. Furusawa, G. Spedalieri, G. Adesso, S.
Mancini, G. Chiribella, O. Hirota, B. Munro, N. Metwally, R. Namiki, P.
Tombesi, S. Guha, M. J. W. Hall, S. Danilishin, R. G. Patron, M. Bellini, R.
Filip, and J. Eschner for discussions, suggestions and comments.

\appendix

\section{Entanglement-Breaking conditions for qubit depolarizing channels
\label{DEPapp}}

Here we show the conditions under which the depolarizing channel of
Eq.~(\ref{qubitDEPO}) becomes an entanglement-breaking channel. This means to
find a specific regime for the probabilities $p_{k}$ characterizing the channel.

First of all note that, for Hilbert spaces of finite dimension $d$, a simple
way to check if a quantum channel $\mathcal{E}$ is entanglement-breaking is to
test it on the maximally-entangled state $\left\vert \psi\right\rangle _{AB}$
of Eq.~(\ref{maxENTstate}). In other words, if $(\mathcal{E}_{A}%
\otimes\mathcal{I}_{B})\left\vert \psi\right\rangle _{AB}\left\langle
\psi\right\vert $ is separable, then $(\mathcal{E}_{A}\otimes\mathcal{I}%
_{B})\rho_{AB}$ is separable for any input state $\rho_{AB}$~\cite{EBchannels}%
. In the case of qubits, we can test the channel on the triplet state
$\left\vert +\right\rangle _{AB}$ of Eq.~(\ref{tripletSTATE}). We then compute
the output state%
\begin{align}
\Phi &  :=(\mathcal{E}_{A}\otimes\mathcal{I}_{B})(\left\vert +\right\rangle
_{AB}\left\langle +\right\vert )\nonumber\\
&  =\sum_{k=0}^{3}p_{k}(P_{k}\otimes I)\left\vert +\right\rangle
_{AB}\left\langle +\right\vert (P_{k}^{\dagger}\otimes I)\nonumber\\
&  =p_{0}\left\vert +\right\rangle _{AB}\left\langle +\right\vert \nonumber\\
&  +p_{1}(X\otimes I)\left\vert +\right\rangle _{AB}\left\langle +\right\vert
(X\otimes I)\\
&  +p_{2}(Y\otimes I)\left\vert +\right\rangle _{AB}\left\langle +\right\vert
(Y^{\dagger}\otimes I)\nonumber\\
&  +p_{3}(Z\otimes I)\left\vert +\right\rangle _{AB}\left\langle +\right\vert
(Z\otimes I)~.
\end{align}
Adopting the computational basis $\{\left\vert 00\right\rangle ,\left\vert
01\right\rangle ,\left\vert 10\right\rangle ,\left\vert 11\right\rangle \}$
and using $X\left\vert u\right\rangle =\left\vert u\oplus1\right\rangle $,
$Z\left\vert u\right\rangle =(-1)^{u}\left\vert u\right\rangle $ and $Y=iXZ$,
we get%
\begin{equation}
\Phi=\sum_{ijkl}\Phi_{ijkl}\left\vert i,j\right\rangle _{AB}\left\langle
k,l\right\vert ~,
\end{equation}
where the coefficients $\Phi_{ijkl}=\left\langle i,j\right\vert \Phi\left\vert
k,l\right\rangle $ are the elements of the following density matrix%
\begin{equation}
\boldsymbol{\Phi}=\frac{1}{2}\left(
\begin{array}
[c]{cccc}%
p_{0}+p_{3} & 0 & 0 & p_{0}-p_{3}\\
0 & p_{1}+p_{2} & p_{1}-p_{2} & 0\\
0 & p_{1}-p_{2} & p_{1}+p_{2} & 0\\
p_{0}-p_{3} & 0 & 0 & p_{0}+p_{3}%
\end{array}
\right)  ~.
\end{equation}
To check the separability properties we adopt the Peres-Horodecki
criterion~\cite{PERES,HOROsep}. This corresponds to compute the partial
transposition (PT) of the state which is given by the following linear map
\begin{align}
\Phi &  =\sum_{ijkl}\Phi_{ijkl}\left\vert i\right\rangle _{A}\left\langle
k\right\vert \otimes\left\vert j\right\rangle _{B}\left\langle l\right\vert
\nonumber\\
&  \rightarrow\mathrm{PT}(\Phi)=\sum_{ijkl}\Phi_{ijkl}\left\vert
i\right\rangle _{A}\left\langle k\right\vert \otimes(\left\vert j\right\rangle
_{B}\left\langle l\right\vert )^{T}\nonumber\\
&  =\sum_{ijkl}\Phi_{ijkl}\left\vert i\right\rangle _{A}\left\langle
k\right\vert \otimes\left\vert l\right\rangle _{B}\left\langle j\right\vert
\nonumber\\
&  =\sum_{ijkl}\Phi_{ilkj}\left\vert i\right\rangle _{A}\left\langle
k\right\vert \otimes\left\vert j\right\rangle _{B}\left\langle l\right\vert ~.
\end{align}
At the level of the density matrix we then have
\begin{equation}
\boldsymbol{\Phi}=((\Phi_{ijkl}))\rightarrow\mathrm{PT}(\boldsymbol{\Phi
})=((\Phi_{ilkj}))~,
\end{equation}
i.e., $j\longleftrightarrow l$ swapping. It is easy to check that the
partially-transpose matrix
\begin{equation}
\mathrm{PT}(\boldsymbol{\Phi})=\frac{1}{2}\left(
\begin{array}
[c]{cccc}%
p_{0}+p_{3} & 0 & 0 & p_{1}-p_{2}\\
0 & p_{1}+p_{2} & p_{0}-p_{3} & 0\\
0 & p_{0}-p_{3} & p_{1}+p_{2} & 0\\
p_{1}-p_{2} & 0 & 0 & p_{0}+p_{3}%
\end{array}
\right)
\end{equation}
has eigenvalues
\begin{equation}
\lambda_{k}=\frac{1}{2}-p_{k}~~(k=0,1,2,3)~.
\end{equation}
Thus the partially-transposed state has the following spectral decomposition%
\begin{equation}
\mathrm{PT}(\Phi)=\sum_{k=0}^{3}\left(  \frac{1}{2}-p_{k}\right)  \left\vert
\eta_{k}\right\rangle \left\langle \eta_{k}\right\vert ~,
\end{equation}
with $\left\vert \eta_{k}\right\rangle $ orthogonal eigenstates. This operator
is positive ($\geq0$) if and only if%
\begin{equation}
p_{k}\leq\frac{1}{2}~,~\text{for}~k=0,1,2,3~.
\end{equation}
As a result, the output state $\Phi$ is separable (and the channel is
entanglement-breaking) if and only if $p_{k}\leq1/2$ for any $k$, as reported
in the main text.

\section{Unitary $2$-design for the $U\otimes U^{\ast}$ twirling channel
\label{TwirlAVEapp}}

Let us consider two qudits with the same dimension $d$, so that the composite
system is described by an Hilbert space $\mathcal{H}=\mathcal{H}_{A}%
\otimes\mathcal{H}_{B}$ with finite dimension $d^{2}$. From the
literature~\cite{Designs,Gross}, we know that we can write the following
equality for any input state $\rho_{AB}$%
\begin{align}
\mathcal{E}_{UU}(\rho_{AB})  &  :=\int_{\mathcal{U}(d)}dU~(U\otimes
U)\rho_{AB}(U\otimes U)^{\dagger}\label{Euu1}\\
&  =\frac{1}{K}\sum_{k=0}^{K-1}(U_{k}\otimes U_{k})\rho_{AB}(U_{k}\otimes
U_{k})^{\dagger}~,\label{Euu2}%
\end{align}
which is valid for $U_{k}\in\mathcal{D}$, where $\mathcal{D}$\ is a unitary
2-design with $K$ elements. Here we can easily show that
\begin{align}
\mathcal{E}_{UU^{\ast}}(\rho_{AB})  &  :=\int_{\mathcal{U}(d)}dU~(U\otimes
U^{\ast})\rho_{AB}(U\otimes U^{\ast})^{\dagger}\label{Euustar1}\\
&  =\frac{1}{K}\sum_{k=0}^{K-1}(U_{k}\otimes U_{k}^{\ast})\rho_{AB}%
(U_{k}\otimes U_{k}^{\ast})^{\dagger}~,\label{Euustar2}%
\end{align}
where $U_{k}$ belongs to the same design $\mathcal{D}$ as before. In a few
words, the two twirling channels, $\mathcal{E}_{UU}$ and $\mathcal{E}%
_{UU^{\ast}}$, can be decomposed using the same unitary 2-design.

In the first step of the proof we show that $\mathcal{E}_{UU}$ and
$\mathcal{E}_{UU^{\ast}}$ are connected by a partial transposition. Consider
an arbitrary input state $\rho_{AB}$ decomposed in the orthonormal basis of
$\mathcal{H}$~\cite{Notation}%
\begin{equation}
\rho_{AB}=\sum_{ijkl}\rho_{ij}^{kl}\left\vert i\right\rangle _{A}\left\langle
j\right\vert \otimes\left\vert k\right\rangle _{B}\left\langle l\right\vert ~.
\end{equation}
Its partial transposition corresponds to the transposition of system $B$ only,
i.e.,%
\begin{equation}
\mathrm{PT}(\rho_{AB})=\sum_{ijkl}\rho_{ij}^{kl}\left\vert i\right\rangle
_{A}\left\langle j\right\vert \otimes\left\vert l\right\rangle _{B}%
\left\langle k\right\vert ~.
\end{equation}
Then, we prove that
\begin{equation}
\mathcal{E}_{UU^{\ast}}(\rho_{AB})=\mathrm{PT}\{\mathcal{E}_{UU}%
[\mathrm{PT}(\rho_{AB})]\}~.\label{PTconnection}%
\end{equation}
In fact, by linearity we have
\begin{gather}
\mathcal{E}_{UU}[\mathrm{PT}(\rho_{AB})]=\int_{\mathcal{U}(d)}dU~(U\otimes
U)\mathrm{PT}(\rho_{AB})(U\otimes U)^{\dagger}\nonumber\\
=\sum_{ijkl}\rho_{ij}^{kl}\int_{\mathcal{U}(d)}dU~U\left\vert i\right\rangle
_{A}\left\langle j\right\vert U^{\dagger}\otimes U\left\vert l\right\rangle
_{B}\left\langle k\right\vert U^{\dagger}.
\end{gather}
Since $[U\left\vert l\right\rangle _{B}\left\langle k\right\vert U^{\dagger
}]^{T}=U^{\ast}\left\vert k\right\rangle _{B}\left\langle l\right\vert
(U^{\ast})^{\dagger}$ we get%
\begin{gather}
\mathrm{PT}\{\mathcal{E}_{UU}[\mathrm{PT}(\rho_{AB})]\}=\nonumber\\
=\sum_{ijkl}\rho_{ij}^{kl}\int_{\mathcal{U}(d)}dU~U\left\vert i\right\rangle
_{A}\left\langle j\right\vert U^{\dagger}\otimes U^{\ast}\left\vert
k\right\rangle _{B}\left\langle l\right\vert (U^{\ast})^{\dagger}\nonumber\\
=\int_{\mathcal{U}(d)}dU(U\otimes U^{\ast})\rho_{AB}(U\otimes U^{\ast
})^{\dagger}=\mathcal{E}_{UU^{\ast}}(\rho_{AB})~.
\end{gather}

Now the second step is to combine Eq.~(\ref{PTconnection}) with the unitary
$2$-design for $\mathcal{E}_{UU}$. First it is important to note that the
equivalence between Eqs.~(\ref{Euu1}) and~(\ref{Euu2}) is valid not only when
$\rho_{AB}$ is a density operator but, more generally, when it is an Hermitian
linear operator. This extension is straightforward to prove. Suppose that the
linear operator $O:\mathcal{H}\rightarrow\mathcal{H}$ is Hermitian. Then its
spectral decomposition involves real eigenvalues and orthonormal eigenvectors,
i.e., we can write%
\begin{equation}
O=\sum_{n}O_{n}\left\vert \phi_{n}\right\rangle \left\langle \phi
_{n}\right\vert ~,
\end{equation}
where $O_{n}\in\mathbb{R}$ and $\left\langle \phi_{n}\right\vert \left.
\phi_{m}\right\rangle =\delta_{nm}$. Now we can write
\begin{gather}
\mathcal{E}_{UU}(O):=\int_{\mathcal{U}(d)}dU~(U\otimes U)O(U\otimes
U)^{\dagger}\label{s1}\\
=\sum_{n}O_{n}\int_{\mathcal{U}(d)}dU~(U\otimes U)\left\vert \phi
_{n}\right\rangle \left\langle \phi_{n}\right\vert (U\otimes U)^{\dagger
}\label{s2}\\
=\sum_{n}O_{n}\frac{1}{K}\sum_{k=0}^{K-1}(U_{k}\otimes U_{k})\left\vert
\phi_{n}\right\rangle \left\langle \phi_{n}\right\vert (U_{k}\otimes
U_{k})^{\dagger}\label{s3}\\
=\frac{1}{K}\sum_{k=0}^{K-1}(U_{k}\otimes U_{k})O(U_{k}\otimes U_{k}%
)^{\dagger}~,\label{s4}%
\end{gather}
where (\ref{s1})$\rightarrow$(\ref{s2}) by linearity, (\ref{s2})$\rightarrow
$(\ref{s3}) by the fact that $\left\vert \phi_{n}\right\rangle \left\langle
\phi_{n}\right\vert $ are projectors (and therefore states) and (\ref{s3}%
)$\rightarrow$(\ref{s4}) by linearity again.

As a result, we can apply the equivalence between Eqs.~(\ref{s1})
and~(\ref{s4})\ to the linear operator $\mathrm{PT}(\rho_{AB})$ which fails to
be a density operator when $\rho_{AB}$ is entangled but still it is Hermitian
(and unit trace) in the general case. Thus, we can write%
\begin{gather}
\mathcal{E}_{UU}[\mathrm{PT}(\rho_{AB})]=\frac{1}{K}\sum_{t=0}^{K-1}%
(U_{t}\otimes U_{t})\mathrm{PT}(\rho_{AB})(U_{t}\otimes U_{t})^{\dagger
}\nonumber\\
=\frac{1}{K}\sum_{t=0}^{K-1}\sum_{ijkl}\rho_{ij}^{kl}U_{t}\left\vert
i\right\rangle _{A}\left\langle j\right\vert U_{t}^{\dagger}\otimes
U_{t}\left\vert l\right\rangle _{B}\left\langle k\right\vert U_{t}^{\dagger}~.
\end{gather}
Now, using the connection in Eq.~(\ref{PTconnection}) and the fact that
$[U_{t}\left\vert l\right\rangle _{B}\left\langle k\right\vert U_{t}^{\dagger
}]^{T}=U_{t}^{\ast}\left\vert k\right\rangle _{B}\left\langle l\right\vert
(U_{t}^{\ast})^{\dagger}$, we can write%
\begin{gather}
\mathcal{E}_{UU^{\ast}}(\rho_{AB})=\mathrm{PT}\{\mathcal{E}_{UU}%
[\mathrm{PT}(\rho_{AB})]\}\nonumber\\
=\frac{1}{K}\sum_{t=0}^{K-1}\sum_{ijkl}\rho_{ij}^{kl}U_{t}\left\vert
i\right\rangle _{A}\left\langle j\right\vert U_{t}^{\dagger}\otimes
U_{t}^{\ast}\left\vert k\right\rangle _{B}\left\langle l\right\vert
(U_{t}^{\ast})^{\dagger}\nonumber\\
=\frac{1}{K}\sum_{t=0}^{K-1}(U_{t}\otimes U_{t}^{\ast})\rho_{AB}(U_{t}\otimes
U_{t}^{\ast})^{\dagger}~,\label{finaPT}%
\end{gather}
which gives the equivalence between Eqs.~(\ref{Euustar1}) and~(\ref{Euustar2}).

\section{Partial Haar average of a linear operator\label{HaarAVEsec}}

In this short appendix we give a simple proof of Eq.~(\ref{depola2}). Consider
an Hilbert space $\mathcal{H}=\mathcal{H}_{A}\otimes\mathcal{H}_{B}$ with
finite dimension $d_{A}d_{B}$ (where $d_{A}$ and $d_{B}$ are generally
different). Given a linear operator $T:\mathcal{H}\rightarrow\mathcal{H}$, we
can always decompose it in an orthonormal basis
\begin{equation}
T=\sum_{ijkl}T_{ij}^{kl}\left\vert i\right\rangle _{A}\left\langle
j\right\vert \otimes\left\vert k\right\rangle _{B}\left\langle l\right\vert ~.
\end{equation}
Then, we can write the following partial Haar average, where only system $A$
is averaged on the unitary group%
\begin{align}
\left\langle T\right\rangle _{U\otimes I}  &  :=\int_{\mathcal{U}%
(d)}dU(U\otimes I)T(U^{\dagger}\otimes I)\nonumber\\
&  =\sum_{ijkl}T_{ij}^{kl}\left(  \int_{\mathcal{U}(d)}dU~U\left\vert
i\right\rangle _{A}\left\langle j\right\vert U^{\dagger}\right)
\otimes\left\vert k\right\rangle _{B}\left\langle l\right\vert .\label{TUI}%
\end{align}
Now we use Eq.~(\ref{twirlingID}) with linear operator $O=\left\vert
i\right\rangle _{A}\left\langle j\right\vert $, which gives%
\begin{equation}
\int_{\mathcal{U}(d)}dU~U\left\vert i\right\rangle _{A}\left\langle
j\right\vert U^{\dagger}=\delta_{ij}\frac{I}{d_{A}}~.
\end{equation}
Then, by replacing this expression in Eq.~(\ref{TUI}), we get%
\begin{align}
\left\langle T\right\rangle _{U\otimes I}  &  =\frac{I}{d_{A}}\otimes
\sum_{ikl}T_{ii}^{kl}\left\vert k\right\rangle _{B}\left\langle l\right\vert
\nonumber\\
&  =\frac{I}{d_{A}}\otimes\mathrm{Tr}_{A}\left(  T\right)  ~.\label{HaarAVE2}%
\end{align}
This is a simple extension of Eq.~(\ref{twirlingID}) to considering the
presence of a second (unaveraged) system $B$. In particular, for $T$ density
operator, we have the result of Eq.~(\ref{depola2}).

\section{Uniformly dephasing channel is
entanglement-breaking\label{DephaseAPP}}

Here we prove that the uniformly dephasing channel of Eq.~(\ref{UNIdepha}) is
entanglement-breaking. First consider a pure input state $\rho_{AB}=\left\vert
\varphi\right\rangle _{AB}\left\langle \varphi\right\vert $ expressed in the
Fock basis of the two modes
\begin{equation}
\left\vert \varphi\right\rangle _{AB}=\sum_{kj}c_{kj}\left\vert k\right\rangle
_{A}\otimes\left\vert j\right\rangle _{B}~,~\sum_{kj}\left\vert c_{kj}%
\right\vert ^{2}=1~.
\end{equation}
Since $\hat{R}_{\theta}\left\vert k\right\rangle =\exp(-i\theta k)\left\vert
k\right\rangle $, we get%
\begin{align}
\rho_{A^{\prime}B}  &  =\sum_{kjk^{\prime}j^{\prime}}c_{kj}c_{k^{\prime
}j^{\prime}}^{\ast}\left(  \int\frac{d\theta}{2\pi}e^{-i\theta(k-k^{\prime}%
)}\left\vert k\right\rangle _{A}\left\langle k^{\prime}\right\vert \right)
\otimes\left\vert j\right\rangle _{B}\left\langle j^{\prime}\right\vert
\nonumber\\
&  =\sum_{kjk^{\prime}j^{\prime}}c_{kj}c_{k^{\prime}j^{\prime}}^{\ast}%
\delta(k-k^{\prime})\left\vert k\right\rangle _{A}\left\langle k\right\vert
\otimes\left\vert j\right\rangle _{B}\left\langle j^{\prime}\right\vert
\nonumber\\
&  =\sum_{kjj^{\prime}}c_{kj}c_{kj^{\prime}}^{\ast}\left\vert k\right\rangle
_{A}\left\langle k\right\vert \otimes\left\vert j\right\rangle _{B}%
\left\langle j^{\prime}\right\vert ~.
\end{align}
Then, by re-distributing the sum, we get%
\begin{align}
\rho_{A^{\prime}B}  &  =\sum_{k}\left(  \sum_{j}c_{kj}\left\vert
kj\right\rangle \right)  \otimes\left(  \sum_{j^{\prime}}c_{kj^{\prime}}%
^{\ast}\left\langle kj^{\prime}\right\vert \right) \nonumber\\
&  =\sum_{k}d_{k}\left(  \sum_{j}\frac{c_{kj}}{\sqrt{d_{k}}}\left\vert
kj\right\rangle \right)  \otimes\left(  \sum_{j^{\prime}}\frac{c_{kj^{\prime}%
}^{\ast}}{\sqrt{d_{k}}}\left\langle kj^{\prime}\right\vert \right)  ~,
\end{align}
where we have introduced $d_{k}:=\sum_{j}\left\vert c_{kj}\right\vert ^{2}$ in
the last step (clearly $\sum_{k}d_{k}=1$). Now introducing the pure state%
\begin{equation}
\left\vert \eta_{k}\right\rangle :=\sum_{j}\frac{c_{kj}}{\sqrt{d_{k}}%
}\left\vert kj\right\rangle ,
\end{equation}
we can write the following spectral decomposition for the output state%
\begin{equation}
\rho_{A^{\prime}B}=\sum_{k}d_{k}\left\vert \eta_{k}\right\rangle \left\langle
\eta_{k}\right\vert ~.
\end{equation}
Now we note that we can always write the tensor product
\begin{equation}
\left\vert \eta_{k}\right\rangle =\left\vert k\right\rangle \otimes\left\vert
\xi(k)\right\rangle ,~\left\vert \xi(k)\right\rangle :=\sum_{j}\frac{c_{kj}%
}{\sqrt{d_{k}}}\left\vert j\right\rangle ,
\end{equation}
so that the output state is manifestly in separable form
\begin{equation}
\rho_{A^{\prime}B}=\sum_{k}d_{k}\left\vert k\right\rangle _{A^{\prime}%
}\left\langle k\right\vert \otimes\left\vert \xi(k)\right\rangle
_{B}\left\langle \xi(k)\right\vert ~.
\end{equation}
Proof can trivially be extended to mixed states via their spectral
decomposition into pure states.

\section{$\hat{R}_{\theta}\otimes\hat{R}_{\theta}$-invariant Gaussian states
are separable\label{invGAUSapp}}

To derive the CM of Eq.~(\ref{CM_appfinite}) just check that a $2\times2$ real
matrix $\mathbf{M}$ is invariant under rotations $\mathbf{R}_{\theta
}\mathbf{MR}_{\theta}^{T}=\mathbf{M}$ (or, equivalently, it commutes with
rotations $[\mathbf{M},\mathbf{R}_{\theta}]=0$) if and only if it takes the
asymmetric form%
\begin{equation}
\mathbf{M}=\left(
\begin{array}
[c]{cc}%
m & t\\
-t & m
\end{array}
\right)  ~,
\end{equation}
with $m,t$ real numbers. Thus, the $\mathbf{A},\mathbf{B},\mathbf{C}$ blocks
of the CM~(\ref{CM_appfinite}) must have this general form, with $\mathbf{A}$
and $\mathbf{B}$ diagonal by the further condition of symmetry.

Then, it is easy to check that CM of Eq.~(\ref{CM_appfinite}) describes a
separable Gaussian state. In fact, using suitable local rotations
$\mathbf{R}_{x}\mathbf{\oplus R}_{y}$ (therefore not changing the separability
properties of the state), we can transform $\mathbf{V}_{AB}$ into the simpler
form%
\begin{equation}
\mathbf{V}_{AB}^{\prime}(\alpha,\beta,\gamma)=\left(
\begin{array}
[c]{cc}%
\alpha\mathbf{I} & \gamma\mathbf{I}\\
\gamma\mathbf{I} & \beta\mathbf{I}%
\end{array}
\right)  ~,\label{VabprimeAPP}%
\end{equation}
where $\gamma=\sqrt{\omega^{2}+\varphi^{2}}$. Without loss of generality,
suppose that $\beta\geq\alpha$ and set $\beta-\alpha:=\delta$. We can always
generate $\rho_{AB}^{\prime}$ with CM\ $\mathbf{V}_{AB}^{\prime}(\alpha
,\beta,\gamma)$ by applying local Gaussian channels $\mathcal{I}_{A}%
\otimes\mathcal{G}_{B}$ to the symmetric Gaussian state $\rho_{AB}%
^{\prime\prime}$ with CM $\mathbf{V}_{AB}^{\prime\prime}=\mathbf{V}%
_{AB}^{\prime}(\alpha,\alpha,\gamma)$. It is sufficient to choose the identity
channel $\mathcal{I}_{A}$ and a Gaussian channel $\mathcal{G}_{B}$ with
additive noise $\delta$ (also known as canonical B2 form~\cite{RMP}). It is
now trivial to check that the state $\rho_{AB}^{\prime\prime}$ is separable.
In fact, $\mathbf{V}_{AB}^{\prime\prime}$ is a \textit{bona-fide} quantum CM
when its parameters $\alpha$ and $\gamma$ satisfy the conditions $\alpha\geq1$
and $|\gamma|\leq\alpha-1$. Then, one can check that the partially-transposed
symplectic eigenvalues of $\mathbf{V}_{AB}^{\prime\prime}$ are greater than
$1$, i.e., the state is separable, when $|\gamma|\leq\sqrt{\alpha^{2}-1}$,
which is a condition always satisfied. Now, since $\rho_{AB}^{\prime\prime}$
is separable, then also $\rho_{AB}^{\prime}$ and $\rho_{AB}$ must be separable
(local operations cannot create entanglement).


\begin{thebibliography}{99}
\bibitem {Notation}Minor remark on matrix notation. We use $\rho_{ijkl}$ when
we refer to kets $\left\vert i,j\right\rangle $ and bras $\left\langle
k,l\right\vert $. We prefer to use $\rho_{ij}^{kl}$ when we distinguish
between system $A$ ($\left\vert i\right\rangle _{A}\left\langle j\right\vert
$) and system $B$ ($\left\vert k\right\rangle _{B}\left\langle l\right\vert
$). Clearly, we have $\rho_{ijkl}=\rho_{ik}^{jl}$.
\end{thebibliography}
\end{document}